\documentclass[epj]{svjour}
\usepackage[utf8]{inputenc}
\usepackage[T1]{fontenc}
\usepackage{float}
\usepackage{amssymb}
\usepackage{amsmath}
\usepackage{hyperref}
\usepackage{placeins}
\usepackage{makeidx}         % allows index generation
\usepackage{graphicx}        % standard LaTeX graphics tool
\usepackage[bottom]{footmisc}% places footnotes at page bottom
\usepackage[table,xcdraw]{xcolor}

\newcommand\GR{{G_{\rm R}}}
\newcommand\GRprime{{\tilde{G}_{\rm R}}}
\newcommand\Pprime{{\tilde{P}}}
\newcommand\Grr{{G_{rr}}}

\newcommand\Gqr{{G_{\rm qr}}}
\newcommand\Gpr{{G_{\rm pr}}}
\newcommand\Gqrd{{G_{\rm qrd}}}
\newcommand\Gqrnd{{G_{\rm qrnd}}}
\newcommand\Wqr{{W_{\rm qr}}}
\newcommand\Wpr{{W_{\rm pr}}}
\newcommand\Wrr{{W_{rr}}}

\begin{document}

\title{Analysis of world terror networks from  the reduced Google matrix of Wikipedia}

\author{
Samer El Zant$^{2}$
Klaus M. Frahm$^{1}$, Katia Jaffr\`es-Runser$^{2}$ and 
Dima L. Shepelyansky$^{1}$}

\institute{
Laboratoire de Physique Th\'eorique du CNRS, IRSAMC, 
Universit\'e de Toulouse, CNRS, UPS, 31062 Toulouse, France
\and
Institut de Recherche en Informatique de Toulouse, Universit\'e de Toulouse, INPT, Toulouse, France
}

\titlerunning{World terror networks from 
the reduced Google matrix of Wikipedia}
\authorrunning{S. El Zant {\it et al.}}

\abstract{We apply the reduced Google matrix method 
to analyse interactions between 95 terrorist groups
and determine their relationships and influence on 64 world countries. 
This is done on the basis of the Google matrix of the English Wikipedia
(2017) composed of 5~416~537 articles which accumulate a great part of global
human knowledge. The reduced Google matrix takes into account the
direct and hidden links between a selection of 159 nodes (articles)
appearing due to all paths of a random surfer moving over the whole network.
As a result we obtain the network structure of terrorist groups
and their relations with selected countries. Using the sensitivity of PageRank
to a weight variation of specific links we determine the geopolitical
sensitivity and influence of specific terrorist groups
on world countries. We argue that this approach can find useful application for
more extensive and detailed data bases analysis.
}


%\date{\today}
\date{Dated: September 7, 2016}

\maketitle

\section{Introduction}
\label{sec:1}

{\it ''A new type of terrorism threatens the world,
driven by networks of fanatics determined to inflict 
maximum civilian and economic damages on distant targets in pursuit of 
their extremist goals''} \cite{sageman1}. The origins of this world wide phenomenon
are under investigation in political, social and religious sciences
(see e.g.  \cite{kepel1,kepel2,sageman1,sageman2} and Refs. therein).
At the same time the number of terrorist groups  is growing in the world \cite{taliban1}
reaching over 100 officially recognized groups acting in various
countries of the world \cite{wikispgroups}. These numbers become quite large and 
the mathematical analysis of multiple interactions between these groups
and their relationships to world countries is getting of great timeliness.
The first steps in this direction are reported in a few publications 
(see e.g. \cite{hicks,latora}) showing that the network science methods 
(see e.g. \cite{dorogovtsev})
should be well adapted to such type of investigations. However, it is 
difficult to obtain a clear network structure with all dependencies which
are emerging from the surrounding world with all its complexity.

In this work we use the approach of the Google matrix $G$ and PageRank algorithm
developed by Brin and Page for large scale WWW network analysis \cite{brin}.
The mathematical and statistical properties of this approach for various networks 
are described in \cite{rmp2015,langville}. The efficiency of these methods
are demonstrated for Wikipedia and world trade networks in 
\cite{eomplos,ermannwtn,lages}. For the analysis of the terror networks
we use the reduced Google matrix approach developed recently 
\cite{frahm,politwiki,geop}. This approach selects from a global large scale network
a subset of nodes of interest and constructs the reduced Google matrix 
$\GR$ for this subset including all indirect links connecting the subset nodes 
via the global network. The analysis of political leaders 
and world countries subsets of Wikipedia networks in various language editions
demonstrated the efficiency of this analysis \cite{politwiki,geop}. Here, for the 
English Wikipedia network (collected in May 2017), we target a subset
of $N_g = 95$ terrorist groups referenced in Wikipedia articles of groups
enlisted as terrorist groups for at least two countries in \cite{wikispgroups} 
(see Table~\ref{tab:groups}). The collection of 24 editions of Wikipedia networks
dated by May 2017 is available at \cite{24wiki2017}.
In addition we select the group of $N_c=64$ related world countries
given in Table~\ref{tab:countries}. This gives us the size of $\GR$ being $N_r=N_g+N_c=159$
that is much smaller then the global Wikipedia network of
$N=5~416~537$ nodes (articles) and $N_\ell = 122~232~932$ links generated by 
quotation links from one article to another. The method of the reduced Google matrix
and the obtained results for interactions 
between terrorist groups and countries  are described in the next Sections.

We note that the analysis  of Wikipedia data and related
networks is now in development  by various groups (see e.g. \cite{gabella,yasseri,rettinger}).
Here we used the matrix methods for analysis of Wikipedia networks.
These methods have their roots at the investigations of
random matrix theory and quantum chaos \cite{guhr}.

\section{Reduced Google matrix}
\label{sec:2}

It is convenient to describe the network of $N$ Wikipedia articles by the Google matrix $G$ constructed from 
the adjacency matrix $A_{ij}$ with elements $1$ if article (node) $j$ 
points to  article (node) $i$ and zero otherwise. 
In this case, elements of the Google matrix take the standard form 
$G_{ij} = \alpha S_{ij} + (1-\alpha) / N$ \cite{brin,rmp2015,langville},
where $S$ is the matrix of Markov transitions with elements  $S_{ij}=A_{ij}/k_{out}(j)$, 
$k_{out}(j)=\sum_{i=1}^{N}A_{ij}\neq0$ being the node $j$ out-degree
(number of outgoing links) and with $S_{ij}=1/N$ if $j$ has no outgoing links (dangling node). 
Here $0< \alpha <1$ is the damping factor  which for a random surfer
determines the probability $(1-\alpha)$ to jump to any node; below we use the standard value $\alpha=0.85$. 
The right eigenvector
of $G$ with the unit eigenvalue gives the PageRank probabilities
$P(j)$ to find a random surfer on a node $j$. We order all nodes by decreasing probability $P$ 
getting them ordered by the PageRank index $K=1,2,...N$ with a maximal probability at $K=1$.
From this global ranking we obtain the local ranking of groups and countries given in 
Tables~\ref{tab:groups},~\ref{tab:countries}.

The reduced Google matrix $\GR$ is constructed for a selected subset of 
nodes (articles) following the method described
in \cite{frahm,politwiki} 
%%% some proposition from Klaus:
and based on concepts of scattering theory 
used in different fields of mesoscopic and nuclear physics or 
quantum chaos \cite{guhr}.
%%% maybe some additional citations behind ``chaos'' in the 
%%% last phrase.
This matrix has $N_r$ nodes and belongs 
to the class of Google matrices. In addition the 
PageRank probabilities of selected $N_r$ nodes are the same 
as for the global network with $N$ nodes,
up to a constant multiplicative factor taking into account that 
the sum of PageRank probabilities over $N_r$
nodes is unity. The matrix $\GR$ is represented as a sum of three matrices 
(components)
$\GR = \Grr + \Gpr + \Gqr$ \cite{politwiki}. 
The first term $\Grr$ is given by the direct links between selected 
$N_r$ nodes in the global $G$ matrix with $N$ nodes, 
the second term $\Gpr$ is rather close to 
the matrix in which each column is given by 
the PageRank vector $P_r$, ensuring that PageRank probabilities of $\GR$ are 
the same as for $G$ (up to a constant multiplier).
Therefore  $\Gpr$ doesn't provide much information about direct 
and indirect links between selected nodes.
The most interesting is the third matrix $\Gqr$ which takes 
into account all indirect links between
selected nodes appearing due to multiple links via 
the global network nodes $N$ \cite{frahm,politwiki}.
The matrix  $\Gqr = \Gqrd + \Gqrnd$ has diagonal ($\Gqrd$)
and nondiagonal ($\Gqrnd$) parts. The part $\Gqrnd$
represents the main interest since it describes indirect interactions between nodes. 
The explicit formulas as well as the mathematical and numerical computation 
methods of all three components of $\GR$ are given 
in \cite{frahm,politwiki,geop}. 

The selected groups and countries are given in Tables~\ref{tab:groups},~\ref{tab:countries}
in order of their PageRank probabilities (given by KG rank column for groups and Rank column for countries, respectively). 
All countries have PageRank probabilities being larger than those of terrorist groups so that they are well separated.

\FloatBarrier

\section{Results}
\label{sec:3}

In this work we extract from $\GR$ a network of 64 countries and 95 groups. 
This network reflects direct and indirect interactions between countries and groups, 
which motivates us to study the relative influence of group alliances 
on the other ones and on the countries. 
The matrix $\GR$ and its three components $\Grr$, $\Gpr$ and $\Gqr$ 
are computed for $N_r=159$ Wikipedia network nodes 
formed by $N_c=64$ country nodes and $N_g=95$ group nodes.  
The weights of these three $\GR$ components are 
$\Wrr$=0.0644, $\Wpr$=0.8769 and $\Wqr$=0.0587 
(the weight is given by the sum of all matrix elements divided by $N_r$,
thus $\Wrr + \Wqr + \Wqr = 1$). 
The dominant component is $\Gpr$ but as stated above it is approximately given by
columns of the PageRank vector so that the most interesting information is provided by
$\Grr$ and especially the component $\Gqr$ given by indirect links \cite{politwiki,geop}.

\begin{figure}[h]
%\sidecaption
\includegraphics[scale=0.195]{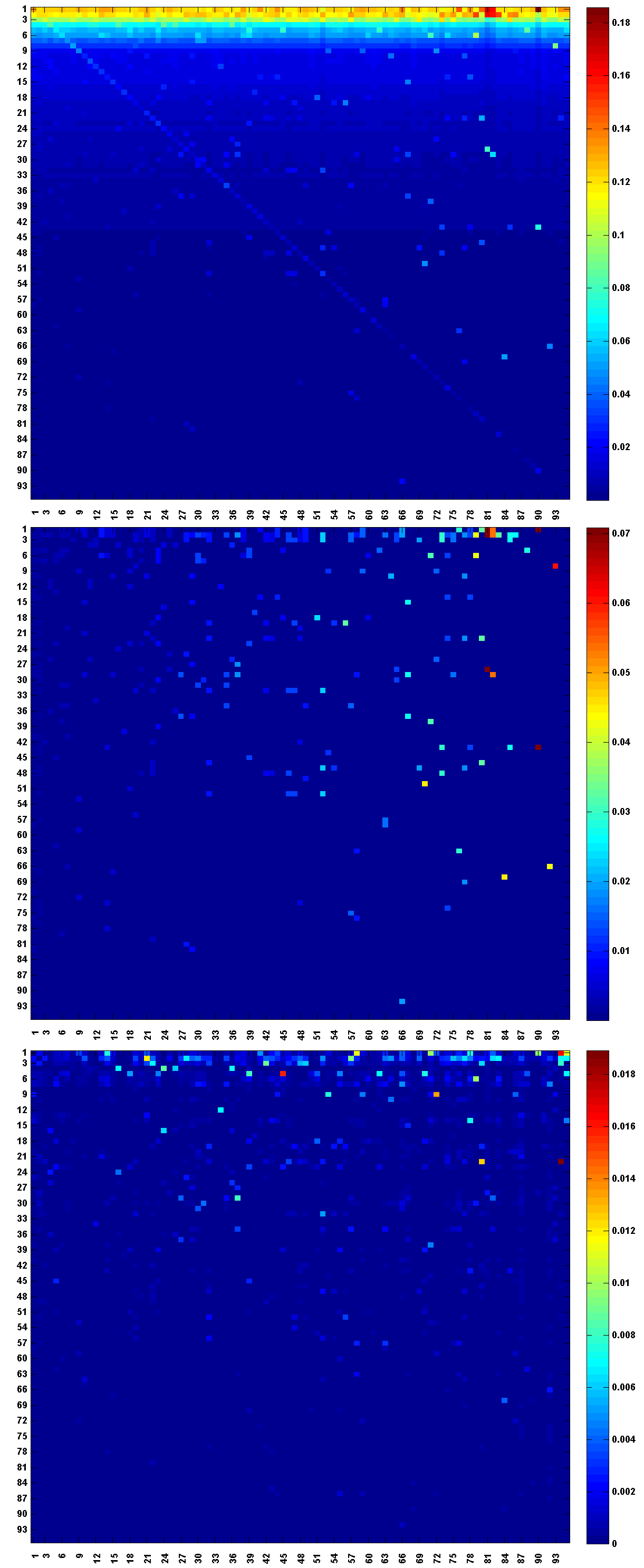}
\caption{Density plots of matrices $\GR$, $\Grr$ and $\Gqrnd$ 
(top, middle  and bottom; color changes from red at maximum to blue at zero); only 95 terrorist nodes of 
Table~\ref{tab:groups} are shown.}
\label{fig1}       % Give a unique label
\end{figure}

The matrix elements of $\GR, \Grr, \Gqr$ corresponding to the part of 
95 terrorist groups are shown in the color maps of Fig.~\ref{fig1} (indices are ordered by increasing 
values of KG as given in Table~\ref{tab:groups}, thus element with KG1=KG1 is located at the top left corner).
The largest matrix elements of $\GR$ are the ones of top PageRank groups of Table~\ref{tab:groups}. 
Such large values are enforced by $\Gpr$ component which is dominated by PageRank vector. 
The elements of $\Grr$ and $\Gqr$ are smaller
but they determine direct and indirect interactions between groups.

According to  Fig.~\ref{fig1} the strong interactions between groups can be found by analyzing $\Gqr$ 
looking at new links appearing in $\Gqr$ and being absent from $\Grr$. As an example we list:
\begin{itemize}
\item[-] Tehrik-i-Taliban Pakistan (KG22) and Jundallah (KG94); 
\item[-] Hamas (KG5) and Izz ad-Din al-Qassam Brigades (KG45);
\item[-] Taliban (KG3) and Al-Qaeda in the Arabian Peninsula (KG21);
 \item[-] Kurdistan Freedom Hawks (KG72) and Kurdistan Workers' Party (KG9). 
\end{itemize}

\subsection{Network structure of groups}
To analyze the network structure of groups we attribute them to 6 different categories marked by 6 colors
in Table~\ref{tab:groups}:  

C1 for the International category of groups operating worldwide 
(color BL, top group is KG1 ISIS) ;  

C2 for the groups targeting Asian countries (color RD, top group is KG3 Taliban) ; 

C3 for the groups related with the Israel-Arab conflict (color OR, top group is KG5 Hamas) ; 

C4 for the groups targeting African countries (color GN, top group is KG10 Al-Shabaab) ; 

C5 for the groups related to Arab countries at Middle East and the Arabian Gulf (color PK, top group is KG13 Houthis) ; 

C6 for all remaining groups (color BK, top group is KG4 IRA). 

These 6 categories of groups
is related to their activity and their geographical location. Only the category C1 has global 
international activity, other categories have more local geographical activity.
We will see that the network analysis captures these categories.

\begin{figure}[h]
%\sidecaption
\includegraphics[scale=0.15]{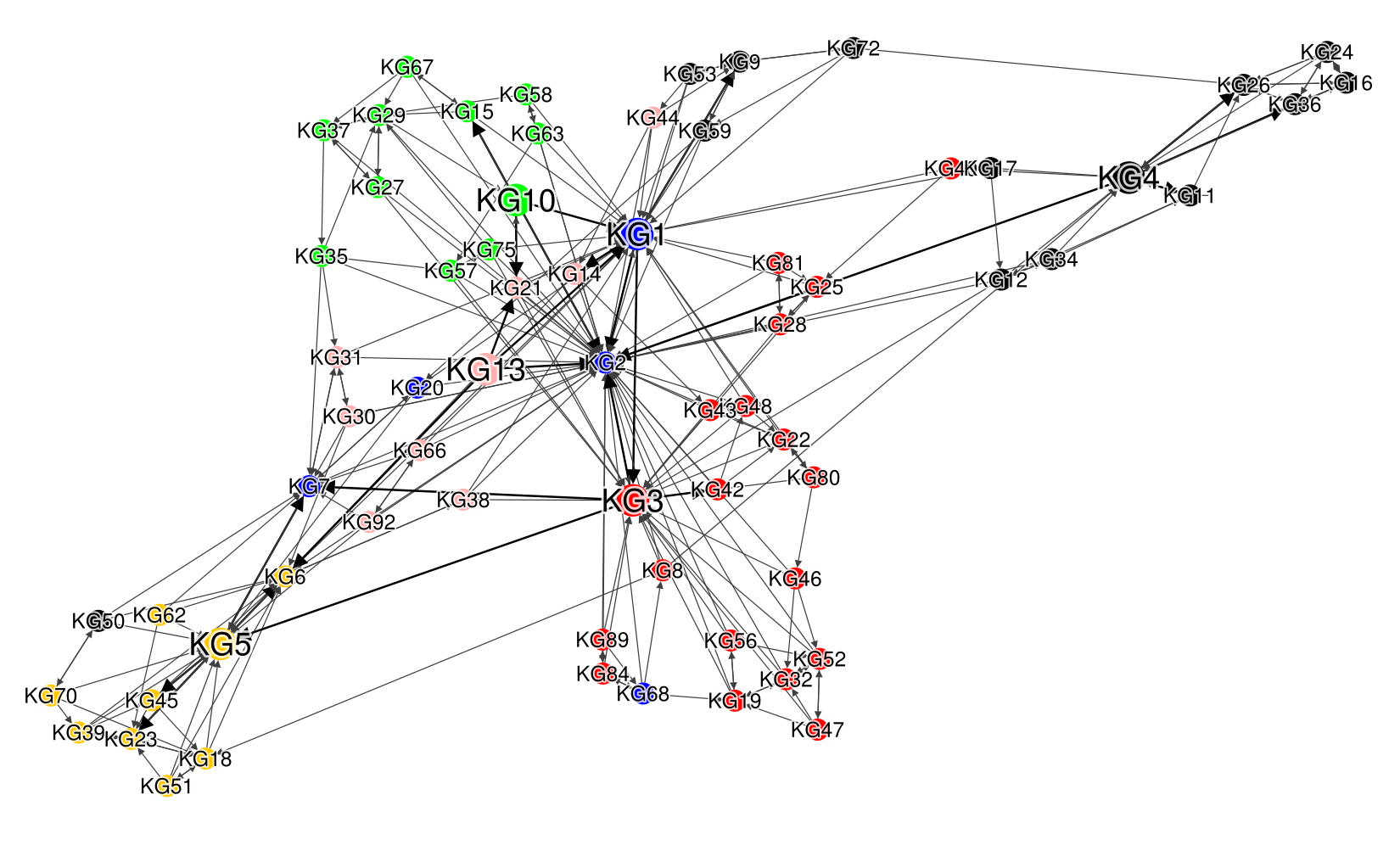}
\caption{Friendship network structure between terrorist groups obtained from $\Gqr$+$\Grr$; colors mark 
categories of nodes and top nodes are given in text and  Table~\ref{tab:groups}; circle size is proportional to PageRank probability of nodes;
bold black arrows point to top 4 friends, gray tiny arrows show friends of friends interactions 
computed until no new edges are added to the graph (drawn with \cite{gephi,hu}.}
\label{fig2}       % Give a unique label
\end{figure}

We analyze the network structure of groups by selecting the top group node of each category 
in Table~\ref{tab:groups} and then, their top 4 friends in $\Grr+\Gqrnd$ 
(i.e. the nodes with the 4 largest matrix elements of $\Grr+\Gqrnd$ 
in the column representing the group of interest. It corresponds to the 4 largest outgoing link weights). 
From the set of top group nodes and their top 4 friends, we continue to extract 
the top 4 friends of friends until no new node is added to this network of friends. 
The obtained network structure of groups is shown in Fig~\ref{fig2}. 
This network structure clearly highlights the clustering of nodes corresponding 
to selected categories. It shows the leading role of top PageRank nodes 
for each category appearing as highly central nodes with large in-degree. 
We note that we speak about networks of friends and followers
using the terminology of social networks. Of course, this has only 
associative meaning (we do not mean that some country is a friend
of terrorist group).

The appearance of links due to indirect relationships between groups is confirmed by well-known facts. 
For instance, it  can be seen that Al-Qaeda in the Arabian Peninsula (KG21) is linking Al-Shabaab (KG10) and Houthis (KG13). Al-Qaeda in the Arabian Peninsula is primarily active in Saudi Arabia. 
It is well known that Saudi Arabia is an important financial support of Al-Shabaab \cite{alshabab1} and that Houthis is confronting Saudi Arabia. As such, it makes sense that Al-Qaeda in the Arabian Peninsula links both groups as it is tied to Saudi Arabia. 

Another meaningful example is the one of Hezbollah (KG6) and Houthis that share the same ideology, since they are both Shiite and are strongly linked to Iran. From Fig.~\ref{fig2}, it can be seen that Hezbollah is a direct friend of Houthis. The case of Hamas (KG5) and Hezbollah, that share the same ideology in facing Israel, is highlighted as well in our results. Moreover, Fig.~\ref{fig2} shows as well that Hezbollah is the linking group between Hamas and Houthis. Finally, the network of Fig.~\ref{fig2} clearly shows that the groups that are listed as International (blue color) are clearly playing that role by having lots of ingoing links from the other categories.

\subsection{Relationships between groups and countries}
\begin{figure}[h]
%\sidecaption
\includegraphics[scale=0.15]{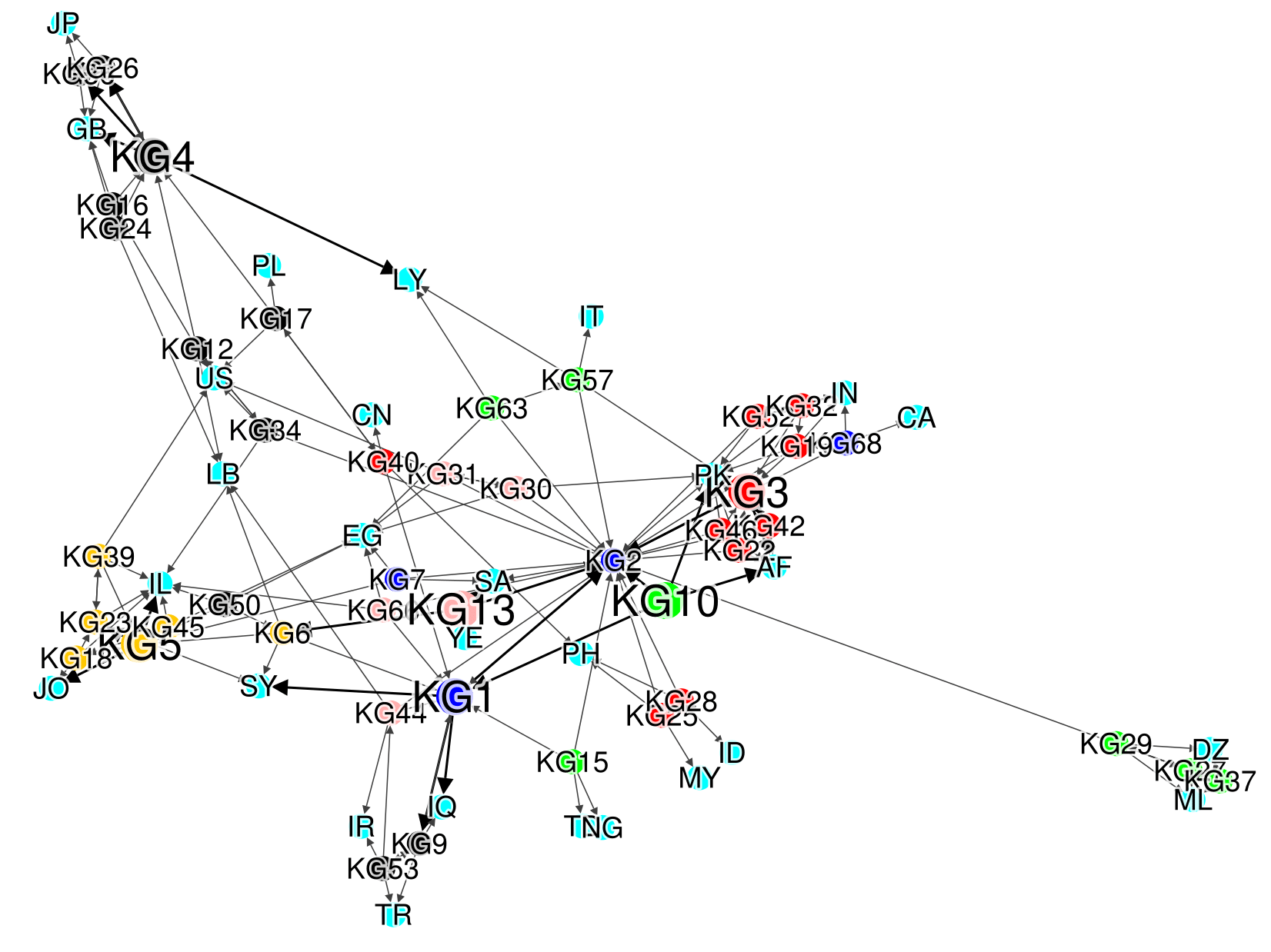}\\
\includegraphics[scale=0.15]{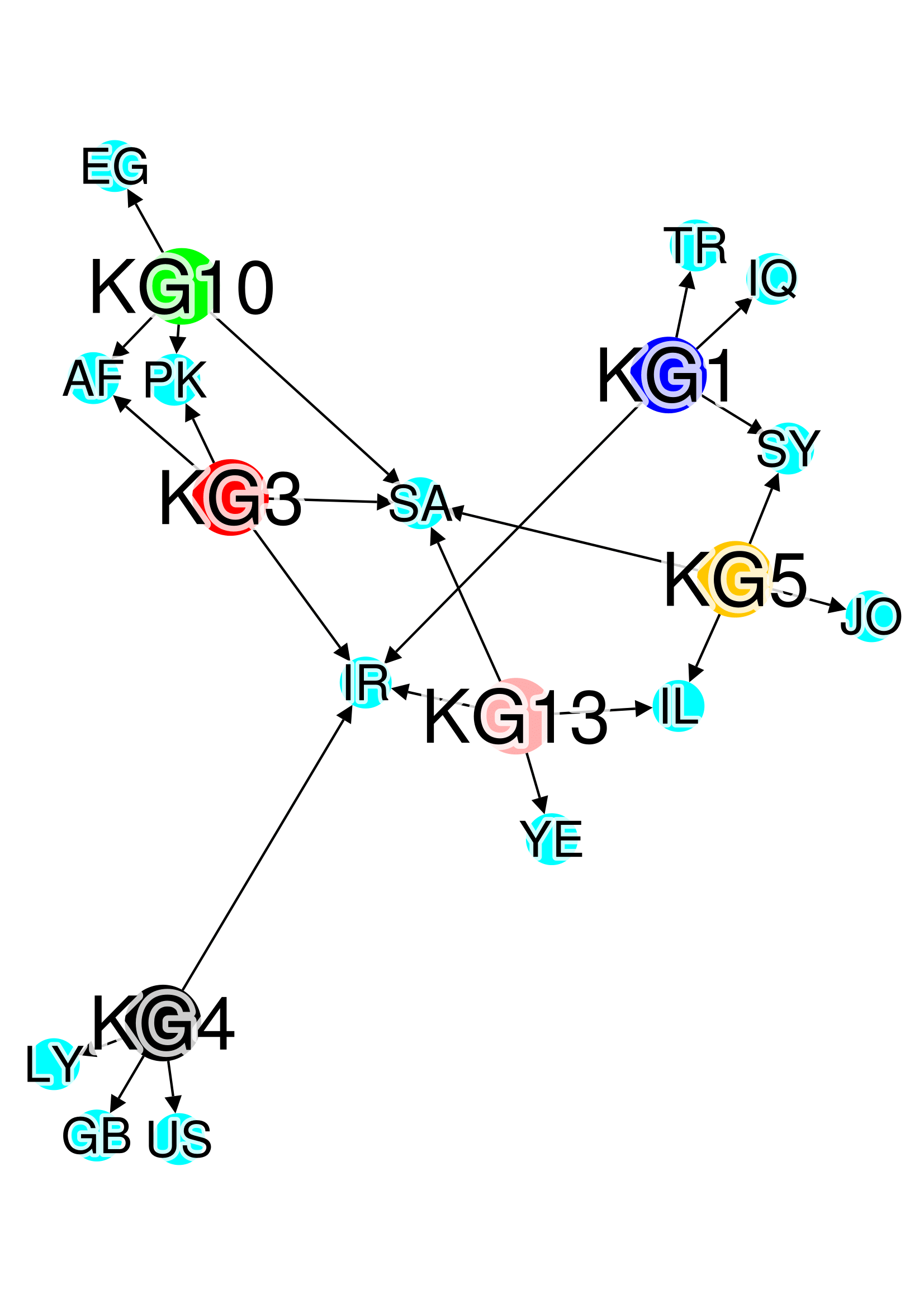}
\caption{Friendship network structure extracted from $\Gqr+\Grr$ 
with the top terrorist groups (marked by their  respective colors) and 
countries (marked by cyan color).
Top panel: the network in case of 2 friends for top terrorist groups
of each category and top friend 2 countries for each group.
Bottom panel: friendship network structure with the top terrorist groups of each category
and their top 4 friend countries.
Networks are drawn with \cite{gephi,hu}.}
\label{fig3}       % Give a unique label
\end{figure}

The interactions between groups and countries are characterized by the network structure
shown in Fig.~\ref{fig3}. For clarity, we first show on the right panel of Fig.~\ref{fig3} the top 4 country friends of the 6 terrorist groups identified as leading each category. On the left panel, we show for the same 6 leading terrorist groups the top 2 country friends and top 2 terrorist groups friends. This latter representation shows altogether major ties between groups and countries and in-between groups. Very interesting and realistic relations between groups and countries can be extracted from this network. For instance, Taliban (KG3) is an active group in Afghanistan and Pakistan that represents an Islamist militant organization that was one of the prominent factions  in the Afghan Civil War \cite{taliban3,taliban2,taliban1}. As shown in Fig.\ref{fig3}, Afghanistan and Pakistan are the countries the most influenced by Taliban.

The fact that Saudi Arabia links Houthis, Taliban and Al Shabaab can be explained 
by the fact that Saudi Arabia is in war with Houthis \cite{yemen1,yemen2}. 
Also, the main funding sources for groups active in Afghanistan and Pakistan 
originate from Saudi Arabia \cite{sauditaliban}. Moreover, Al-Shabaab advocates 
for the Saudi-inspired Wahhabi version of Islam \cite{alshabab2}.
Referring to \cite{isis}, ISIS (KG1) was born in 2006 in Iraq as Islamic State of Iraq (ISI). 
Its main activities are in Syria and Iraq. As shown in Fig.~\ref{fig3} 
a strong relationship exists among the two countries and ISIS.
Hamas and Hezbollah are the leading groups in MEA facing Israel. 
As shown in Fig.\ref{fig3} and knowing the relationship between Hezbollah and Houthis, 
we can explain why Israel is a linking node between Houthis and Hamas.
Finally, we find that Iran links Houthis with ISIS. 
This could be explained by the fact that both groups are in conflict with Saudi Arabia.

\subsection{Sensitivity analysis}
To analyze more specifically the influence of given terrorist groups on
the selected 64 world countries we introduce the sensitivity $F$
determined by the logarithmic derivatives of PageRank probability $P$ 
obtained from $\GR$. At first we define $\delta_{ij}$ as the relative fraction 
to be added to the relationship from node $j$ to node $i$ in $\GR$. 
Knowing $\delta_{ij}$, a new modified matrix $\GRprime$ is calculated in two steps. 
First, element $\GRprime(i,j)$ is set to $(1+\delta_{ij})\cdot\GR(i,j)$. 
Second, all elements of column $j$ of $\GRprime$ are normalized to 1 
(including element $i$) to preserve the column-normalized property of this matrix
from the class of Google matrices.
After that $\GRprime$ reflects an increased probability for going from node $j$ to node $i$.

It is now possible to calculate the modified 
PageRank eigenvector $\Pprime$ from $\GRprime$ using the standard 
$\GRprime\Pprime = \Pprime$ relation and compare it to 
the original PageRank probabilities $P$ calculated with $\GR$ using $\GR P = P$.
Due to the relative change of the transition probability between nodes $i$ and $j$, 
the steady state PageRank probabilities are modified. This reflects 
a structural modification of the network and 
entails a change of importance of nodes in the network. 
These changes are measured by a logarithmic derivative of the PageRank probabilities:

\begin{equation}
%D_{(j \rightarrow i)}(k) = -{({\rm d}P_k}/{d \delta_{ij})/P_k} = -{(\Pprime_k - P_k)}/({\delta_{ij}}{P_k}) 
D_{(j \rightarrow i)}(k) = {({\rm d}P_k}/{{\rm d} \delta_{ij})/P_k} = {(\Pprime_k - P_k)}/({\delta_{ij}}{P_k}) 
\label{eq_sensitivity}
\end{equation}
so that the derivative $D_{(j \rightarrow i)}(k)$ %is a vector and $D_{(j \rightarrow i)}(k)$ 
gives for node $k$ its sensitivity to the change of link $j$ to $i$. 
We note that this approach is similar to the sensitivity analysis 
of the world trade network to the price of specific products 
(e.g. gas or petroleum) as studied in \cite{ermannwtn}.

\begin{figure}[h]
%\sidecaption
\includegraphics[scale=0.07]{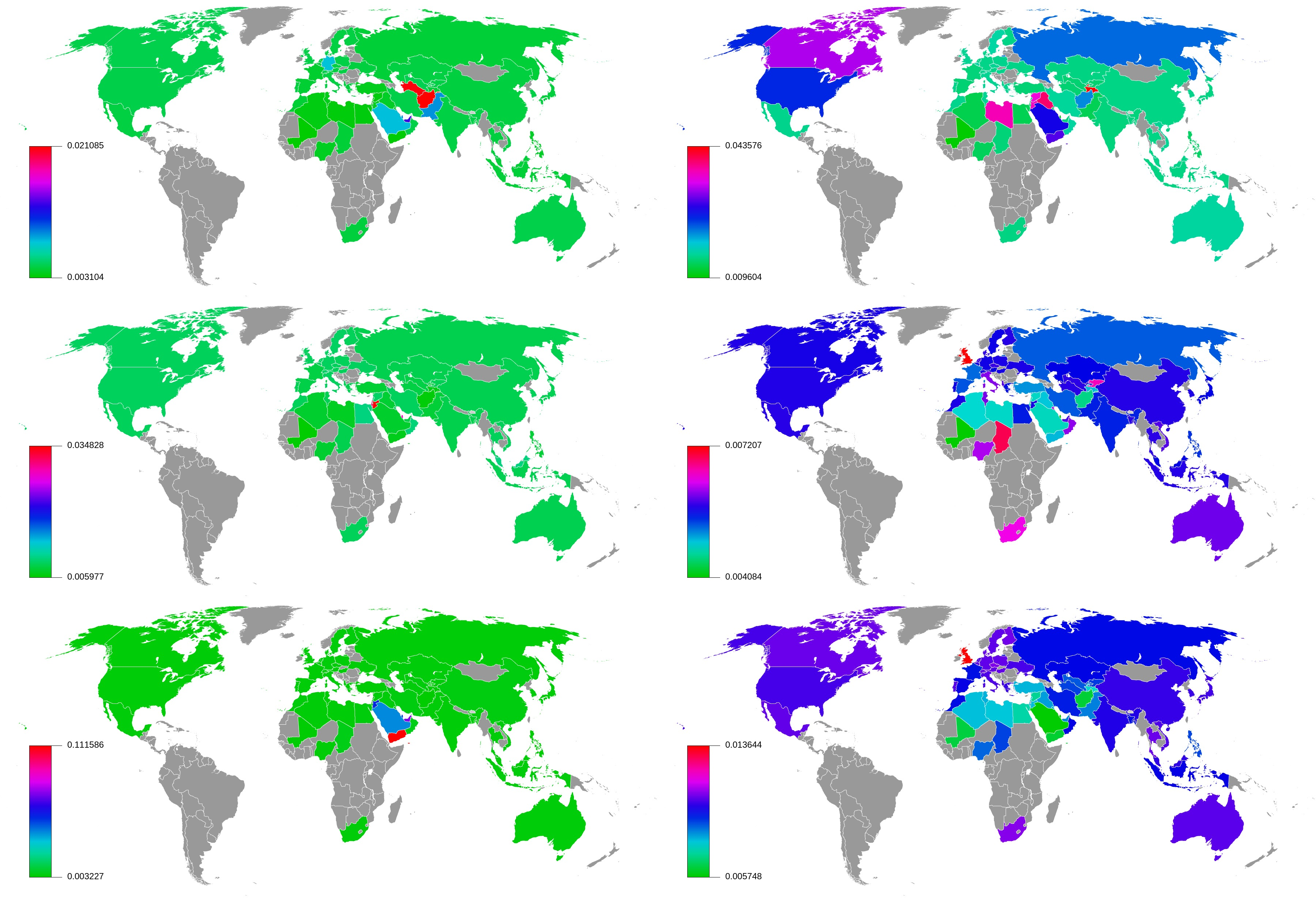}
%
% If no graphics program available, insert a blank space i.e. use
%\picplace{5cm}{2cm} % Give the correct figure height and width in cm
%
\caption{Wold map of the influence of terrorist groups on countries expressed by
sensitivity $D_{(j \rightarrow i)}(j)$ where $j$ 
is the country index and $i$ the group index, see text). Left column:
Taliban KG3, Hamas KG5,  Houthis KG13 (top to bottom). Right column:
ISIS KG1, Al Shabaab KG10, IRA KG4 (top to bottom). Color bar 
marks  $D_{(j \rightarrow i)}(j)$ values with red for maximum and green for minimum influence;
grey color marks countries not considered is this work.}
\label{fig4}       % Give a unique label
\end{figure}

Fig.~\ref{fig4} shows maps of the sensitivity influence $D$ of the top groups 
of the 6 categories on all 64 countries. Here we see that Taliban (KG3) 
has important influence on Afghanistan,  Pakistan, and Saudi Arabia
and less influence on other countries. In contrast ISIS (KG1) 
has a strong worldwide influence with the main effects on
Canada, Libya, USA, Saudi Arabia.
The world maps show that the groups of the left column (Taliban, Hamas, Houthis)
produce mainly local influence in the world.
In contrast, the groups of the right column (ISIS, Al Shabaab, IRA)
spread their influence worldwide. Even if IRA mainly affects UK
it still spreads its influence on other Anglo-Saxon countries.
The presented results determine 
the geopolitical influence of each terrorist group.

\begin{figure}[t]
%\sidecaption
\includegraphics[scale=0.22]{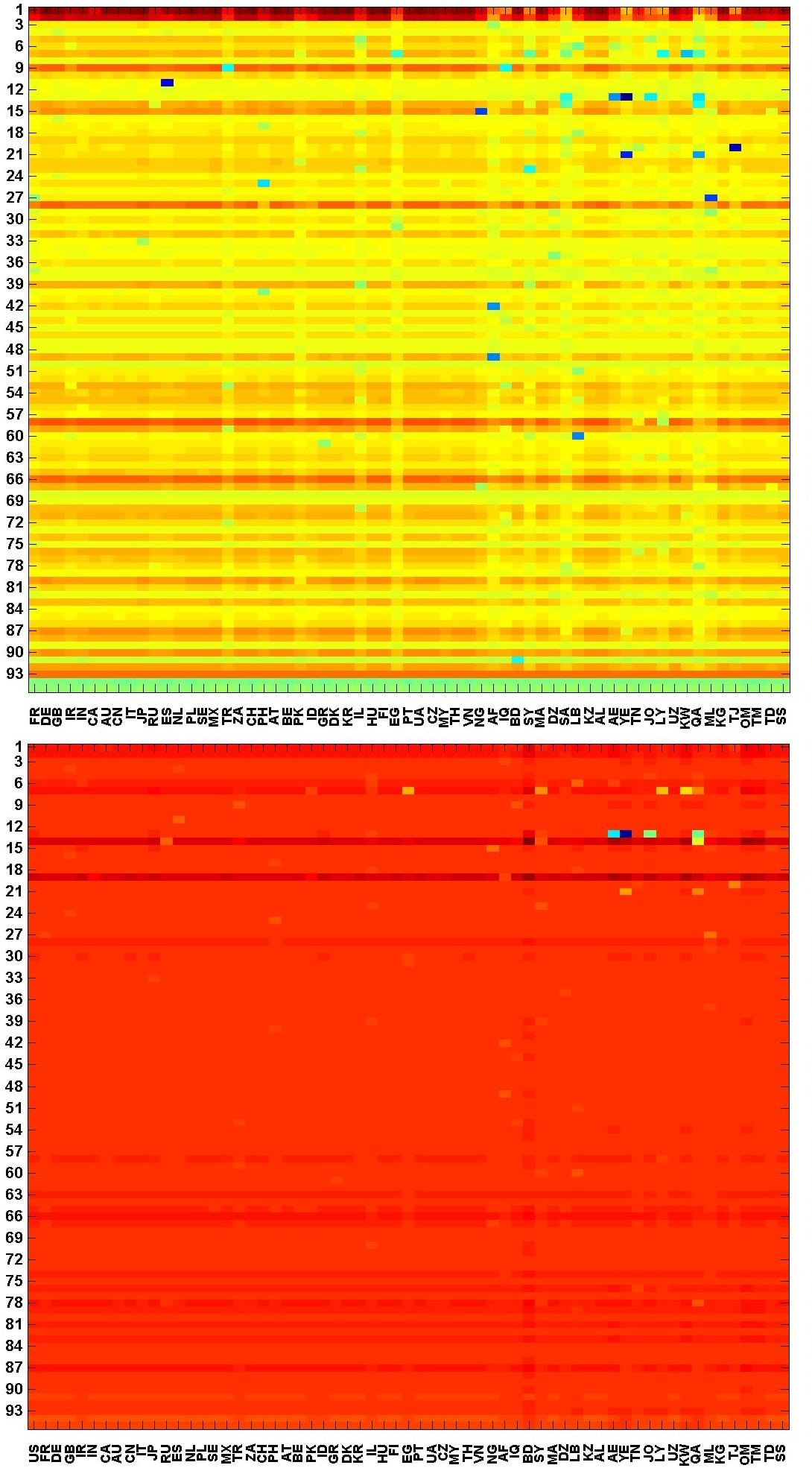}
%
% If no graphics program available, insert a blank space i.e. use
%\picplace{5cm}{2cm} % Give the correct figure height and width in cm
%
\caption{Sensitivity influence $D_{(c \rightarrow i)}(j)$ 
for the relation between a selected country $c$ and a terrorist group $i$
(represented by group index $i$ from Table~\ref{tab:groups} in vertical axis) 
on a world country $j$ (represented by country index $j$ from Table~\ref{tab:countries}
in horizontal axis, $j = c$ is excluded) for two $c$ values: USA (top), 
Saudi Arabia (bottom). Color shows
$D_{(c \rightarrow i)}(j)$ value is changing in 
the range $(-2.8 \cdot 10^{-4},2.1 \cdot 10^{-4})$) for USA
and  $(-4.8 \cdot 10^{-3}, 10^{-3})$) for SA; 
minimum/maximum values correspond to blue/red.}
\label{fig5}       % Give a unique label
\end{figure}

Fig.~\ref{fig5} shows the influence of a relation between one selected country $c$ 
and one selected terrorist group $i$ on the other countries $j$.  
The results are shown for two countries being US (left panel - $c=1$) and 
Saudi Arabia (right panel - $c=46$). Each element $(i,j)$ of the given matrices 
is expressed by $D_{(c \rightarrow i)}(j)$). 
Results show the enormous influence of Saudi Arabia
on terrorist groups and other countries (almost all panel is in red). 
The influence of USA is more selective.

All data for the matrices discussed above, figures and
sensitivity  are available at \cite{wikispgroups}.

\section{Discussion}
\label{sec:4}

We have applied the reduced Google matrix analysis 
(Fig.~\ref{fig1}) to the network of articles of English Wikipedia to analyze the
 network structure of 95 terrorist groups and their influence over 64 world countries
(159 selected articles).
This approach takes into account all human knowledge  accumulated in Wikipedia, 
leveraging all indirect interactions existing between the 159 selected articles and the huge
information contained by  5~416~537 articles of Wikipedia and its 122~232~932 links.
The network structure obtained for the terrorist groups (Figs.~\ref{fig2},~\ref{fig3})
clearly show the presence of 6 types (categories)
of groups. The main groups in each category are determined from their PageRank.
We show that the indirect or hidden links between terrorist groups and countries
play an important role and are, in many cases, predominant over direct links. 
The geopolitical influence of specific terrorist groups on world countries 
is determined via the sensitivity of PageRank variation in respect to specific
links between groups and countries (Fig.~\ref{fig4}).
We see the presence of terrorist groups with localized
geographical influence (e.g. Taliban) and others with worldwide influence (ISIS).
The influence of selected countries on terrorist groups and other countries
is also determined by the developed approach (Fig.6).
The obtained results, tested on the publicly available data of Wikipedia,
show the efficiency of the analysis.
We argue that the reduced Google matrix approach can find further important
applications for terror networks analysis using more advanced and detailed databases.

\section{Acknowledgments}
We thank Sabastiano Vigna \cite{vigna} for providing us his computer codes
which we used for a generation of the English Wikipedia network (2017).
These codes had been 
developed in the frame of EC FET Open NADINE project (2012-2015) \cite{nadine}
and used for Wikipedia (2013) data in \cite{eomplos}.
This work was granted access to the HPC resources of 
CALMIP (Toulouse) under the allocation 2017-P0110. 
This work has been supported by the \emph{GOMOBILE} 
project supported jointly by 
University of Toulouse APR 2015 and R\'egion Occitanie 
under doctoral research grant \#15050459, 
and in part by CHIST-ERA MACACO project, ANR-13-CHR2-0002-06.
This work was granted access to the HPC resources of 
CALMIP (Toulouse) under the allocation of 2017. 

%\vskip -1.5cm

\onecolumn

% \usepackage[normalem]{ulem}
\begin{tiny}
\begin{table}[]
\centering
\caption{List of selected terrorist groups (from \cite{wikispgroups}) attributed to 6 categories marked by color,
$KG$ gives the local PageRank index of terrorist groups.}
\label{tab:groups}
%\begin{tabular}{|c|c|c|c|c|c|}
\begin{tabular}{|p{5cm}|c|c|p{5cm}|c|c|}
\hline
Name                                          & KG  & Color                     & Name                                                    & KG  & Color                     \\ \hline
Islamic State of Iraq and the Levant          & 1  & {\color[HTML]{3531FF} BL} & Hezb-e Islami Gulbuddin                                 & 49 & {\color[HTML]{FE0000} RD} \\ \hline
Al-Qaeda                                      & 2  & {\color[HTML]{3531FF} BL} & Kach and Kahane Chai                                    & 50 & BK                        \\ \hline
Taliban                                       & 3  & {\color[HTML]{FE0000} RD} & Palestine Liberation Front                              & 51 & {\color[HTML]{F56B00} OR} \\ \hline
Provisional Irish Republican Army             & 4  & BK                        & Harkat-ul-Mujahideen                                    & 52 & {\color[HTML]{FE0000} RD} \\ \hline
Hamas                                         & 5  & {\color[HTML]{F56B00} OR} & Kurdistan Free Life Party                               & 53 & BK                        \\ \hline
Hezbollah                                     & 6  & {\color[HTML]{F56B00} OR} & Indian Mujahideen                                       & 54 & {\color[HTML]{FE0000} RD} \\ \hline
Muslim Brotherhood                            & 7  & {\color[HTML]{3531FF} BL} & Abu Nidal Organization                                  & 55 & {\color[HTML]{F56B00} OR} \\ \hline
Liberation Tigers of Tamil Eelam              & 8  & {\color[HTML]{FE0000} RD} & Hizbul Mujahideen                                       & 56 & {\color[HTML]{FE0000} RD} \\ \hline
Kurdistan Workers' Party                      & 9  & BK                        & Libyan Islamic Fighting Group                           & 57 & {\color[HTML]{009901} GN} \\ \hline
Al-Shabaab (militant group)                   & 10 & {\color[HTML]{009901} GN} & Islamic State of Iraq and the Levant in Libya           & 58 & {\color[HTML]{009901} GN} \\ \hline
ETA (separatist group)                        & 11 & BK                        & Revolutionary People's Liberation Party/Front           & 59 & BK                        \\ \hline
FARC                                          & 12 & BK                        & Al-Mourabitoun                                          & 60 & {\color[HTML]{009901} GN} \\ \hline
Houthis                                       & 13 & {\color[HTML]{FF00FF} PK} & Revolutionary Organization 17 November                  & 61 & BK                        \\ \hline
Al-Nusra Front                                & 14 & {\color[HTML]{FF00FF} PK} & Holy Land Foundation for Relief and Development         & 62 & {\color[HTML]{F56B00} OR} \\ \hline
Boko Haram                                    & 15 & {\color[HTML]{009901} GN} & Ansar al-Sharia (Libya)                                 & 63 & {\color[HTML]{009901} GN} \\ \hline
Ulster Volunteer Force                        & 16 & BK                        & Al-Itihaad al-Islamiya                                  & 64 & {\color[HTML]{009901} GN} \\ \hline
Shining Path                                  & 17 & BK                        & Al-Haramain Foundation                                  & 65 & {\color[HTML]{3531FF} BL} \\ \hline
Popular Front for the Liberation of Palestine & 18 & {\color[HTML]{F56B00} OR} & Ansar Bait al-Maqdis                                    & 66 & {\color[HTML]{FF00FF} PK} \\ \hline
Lashkar-e-Taiba                               & 19 & {\color[HTML]{FE0000} RD} & Ansaru                                                  & 67 & {\color[HTML]{009901} GN} \\ \hline
Hizb ut-Tahrir                                & 20 & {\color[HTML]{3531FF} BL} & Babbar Khalsa                                           & 68 & {\color[HTML]{3531FF} BL} \\ \hline
Al-Qaeda in the Arabian Peninsula             & 21 & {\color[HTML]{FF00FF} PK} & Jamaat-ul-Mujahideen Bangladesh                         & 69 & {\color[HTML]{FE0000} RD} \\ \hline
Tehrik-i-Taliban Pakistan                     & 22 & {\color[HTML]{FE0000} RD} & Force 17                                                & 70 & {\color[HTML]{F56B00} OR} \\ \hline
Islamic Jihad Mov. in Palestine           & 23 & {\color[HTML]{F56B00} OR} & Kata'ib Hezbollah                                       & 71 & {\color[HTML]{FF00FF} PK} \\ \hline
%Islamic Jihad Movement in Palestine           & KG23 & {\color[HTML]{F56B00} OR} & Kata'ib Hezbollah                                       & KG71 & {\color[HTML]{FF00FF} PK} \\ \hline
Ulster Defence Association                    & 24 & BK                        & Kurdistan Freedom Hawks                                 & 72 & BK                        \\ \hline
Abu Sayyaf                                    & 25 & {\color[HTML]{FE0000} RD} & Islamic Jihad Union                                     & 73 & {\color[HTML]{FE0000} RD} \\ \hline
Real Irish Republican Army                    & 26 & BK                        & Abdullah Azzam Brigades                                 & 74 & {\color[HTML]{FF00FF} PK} \\ \hline
Ansar Dine                                    & 27 & {\color[HTML]{009901} GN} & Moroccan Islamic Comb. Group                        & 75 & {\color[HTML]{009901} GN} \\ \hline
%Ansar Dine                                    & KG27 & {\color[HTML]{009901} GN} & Moroccan Islamic Combatant Group                        & KG75 & {\color[HTML]{009901} GN} \\ \hline
Jemaah Islamiyah                              & 28 & {\color[HTML]{FE0000} RD} & Ansar al-Sharia (Tunisia)                               & 76 & {\color[HTML]{009901} GN} \\ \hline
Al-Qaeda in the Islamic Maghreb               & 29 & {\color[HTML]{009901} GN} & Al-Qaeda, Indian Subcontinent                     & 77 & {\color[HTML]{FE0000} RD} \\ \hline
%Al-Qaeda in the Islamic Maghreb               & KG29 & {\color[HTML]{009901} GN} & Al-Qaeda in the Indian Subcontinent                     & KG77 & {\color[HTML]{FE0000} RD} \\ \hline
Egyptian Islamic Jihad                        & 30 & {\color[HTML]{FF00FF} PK} & Jund al-Aqsa                                            & 78 & {\color[HTML]{FF00FF} PK} \\ \hline
Al-Jama'a al-Islamiyya                        & 31 & {\color[HTML]{FF00FF} PK} & Hezbollah Al-Hejaz                                      & 79 & {\color[HTML]{FF00FF} PK} \\ \hline
Jaish-e-Mohammed                              & 32 & {\color[HTML]{FE0000} RD} & Jamaat-ul-Ahrar                                         & 80 & {\color[HTML]{FE0000} RD} \\ \hline
Aum Shinrikyo                                 & 33 & {\color[HTML]{FE0000} RD} & Jamaah Ansharut Tauhid                                  & 81 & {\color[HTML]{FE0000} RD} \\ \hline
United Self-Defense Forces of Colombia        & 34 & BK                        & Islamic State of Iraq and the Levant – Algeria Province & 82 & {\color[HTML]{009901} GN} \\ \hline
Armed Islamic Group of Algeria                & 35 & {\color[HTML]{009901} GN} & Osbat al-Ansar                                          & 83 & {\color[HTML]{FF00FF} PK} \\ \hline
Continuity Irish Republican Army              & 36 & BK                        & International Sikh Youth Federation                     & 84 & {\color[HTML]{FE0000} RD} \\ \hline
Movement for Oneness and Jihad in West Africa & 37 & {\color[HTML]{009901} GN} & East Turkestan Liberation Organization                  & 85 & {\color[HTML]{FE0000} RD} \\ \hline
Quds Force                                    & 38 & {\color[HTML]{FF00FF} PK} & Great Eastern Islamic Raiders' Front                    & 86 & BK                        \\ \hline
Al-Aqsa Martyrs' Brigades                     & 39 & {\color[HTML]{F56B00} OR} & Aden-Abyan Islamic Army                                 & 87 & {\color[HTML]{FF00FF} PK} \\ \hline
Com. Party of the Philippines            & 40 & {\color[HTML]{FE0000} RD} & Al-Aqsa Foundation                                      & 88 & {\color[HTML]{F56B00} OR} \\ \hline
%Communist Party of the Philippines            & KG40 & {\color[HTML]{FE0000} RD} & Al-Aqsa Foundation                                      & KG88 & {\color[HTML]{F56B00} OR} \\ \hline
Caucasus Emirate                              & 41 & {\color[HTML]{FE0000} RD} & Khalistan Zindabad Force                                & 89 & {\color[HTML]{FE0000} RD} \\ \hline
Haqqani network                               & 42 & {\color[HTML]{FE0000} RD} & Mujahidin Indonesia Timur                               & 90 & {\color[HTML]{FE0000} RD} \\ \hline
Turkistan Islamic Party                       & 43 & {\color[HTML]{FE0000} RD} & Al-Badr                                                 & 91 & {\color[HTML]{FE0000} RD} \\ \hline
Ansar al-Islam                                & 44 & {\color[HTML]{FF00FF} PK} & Soldiers of Egypt                                       & 92 & {\color[HTML]{FF00FF} PK} \\ \hline
Izz ad-Din al-Qassam Brigades                 & 45 & {\color[HTML]{F56B00} OR} & National Liberation Army                                & 93 & BK                        \\ \hline
Lashkar-e-Jhangvi                             & 46 & {\color[HTML]{FE0000} RD} & Jundallah                                               & 94 & {\color[HTML]{FE0000} RD} \\ \hline
Harkat-ul-Jihad al-Islami                     & 47 & {\color[HTML]{FE0000} RD} & Army of Islam                                           & 95 & {\color[HTML]{FF00FF} PK} \\ \hline
Islamic Movement of Uzbekistan                & 48 & {\color[HTML]{FE0000} RD} &                                                         &      &                           \\ \hline
\end{tabular}
\end{table}
\end{tiny}

\begin{table}[]
\centering
\caption{List of selected countries.}
\label{tab:countries}
\begin{tabular}{|c|c|c|c|c|c|}
\hline
Rank & Name           & abr & Rank & Name                 & abr \\ \hline
1    & United States  & US  & 33   & Portugal             & PT  \\ \hline
2    & France         & FR  & 34   & Ukraine              & UA  \\ \hline
3    & Germany        & DE  & 35   & Czech Republic       & CZ  \\ \hline
4    & United Kingdom & GB  & 36   & Malaysia             & MY  \\ \hline
5    & Iran           & IR  & 37   & Thailand             & TH  \\ \hline
6    & India          & IN  & 38   & Vietnam              & VN  \\ \hline
7    & Canada         & CA  & 39   & Nigeria              & NG  \\ \hline
8    & Australia      & AU  & 40   & Afghanistan          & AF  \\ \hline
9    & China          & CN  & 41   & Iraq                 & IQ  \\ \hline
10   & Italy          & IT  & 42   & Bangladesh           & BD  \\ \hline
11   & Japan          & JP  & 43   & Syria                & SY  \\ \hline
12   & Russia         & RU  & 44   & Morocco              & MA  \\ \hline
13   & Spain          & ES  & 45   & Algeria              & DZ  \\ \hline
14   & Netherlands    & NL  & 46   & Saudi Arabia         & SA  \\ \hline
15   & Poland         & PL  & 47   & Lebanon              & LB  \\ \hline
16   & Sweden         & SE  & 48   & Kazakhstan           & KZ  \\ \hline
17   & Mexico         & MX  & 49   & Albania              & AL  \\ \hline
18   & Turkey         & TR  & 50   & United Arab Emirates & AE  \\ \hline
19   & South Africa   & ZA  & 51   & Yemen                & YE  \\ \hline
20   & Switzerland    & CH  & 52   & Tunisia              & TN  \\ \hline
21   & Philippines    & PH  & 53   & Jordan               & JO  \\ \hline
22   & Austria        & AT  & 54   & Libya                & LY  \\ \hline
23   & Belgium        & BE  & 55   & Uzbekistan           & UZ  \\ \hline
24   & Pakistan       & PK  & 56   & Kuwait               & KW  \\ \hline
25   & Indonesia      & ID  & 57   & Qatar                & QA  \\ \hline
26   & Greece         & GR  & 58   & Mali                 & ML  \\ \hline
27   & Denmark        & DK  & 59   & Kyrgyzstan           & KG  \\ \hline
28   & South Korea    & KR  & 60   & Tajikistan           & TJ  \\ \hline
29   & Israel         & IL  & 61   & Oman                 & OM  \\ \hline
30   & Hungary        & HU  & 62   & Turkmenistan         & TM  \\ \hline
31   & Finland        & FI  & 63   & Chad                 & TD  \\ \hline
32   & Egypt          & EG  & 64   & South Sudan          & SS  \\ \hline
\end{tabular}
\end{table}

\end{document}